\documentclass[letterpaper]{article}

\usepackage{aaai}
\usepackage{times}
\usepackage{helvet}
\usepackage{courier}
\usepackage{url}
\usepackage{graphicx}
\newcommand*{\thead}[1]{\multicolumn{1}{c}{\bfseries #1}}

\usepackage{color}

\title{Pommerman: A Multi-Agent Playground}
\author{
Cinjon Resnick\thanks{Correspondence: cinjon@nyu.edu} \\
NYU\\
\And 
Wes Eldridge \\
Rebellious Labs \\
\And 
David Ha \\
Google Brain \\
\And 
Denny Britz \\
Stanford University \\
\And 
Jakob Foerster \\
University of Oxford \\
\AND
Julian Togelius \\
NYU \\
\And 
Kyunghyun Cho \and Joan Bruna \\
NYU, FAIR
}

\begin{document}
\maketitle


\begin{abstract}
We present Pommerman, a multi-agent environment based on the classic console game Bomberman. Pommerman consists of a set of scenarios, each having at least four players and containing both cooperative and competitive aspects.
We believe that success in Pommerman will require a diverse set of tools and methods, including planning, opponent/teammate modeling, game theory, and communication, and consequently can serve well as a multi-agent benchmark.
To date, we have already hosted one competition, and our next one will be featured in the NIPS 2018 competition track.

\end{abstract}

\section{Why Pommerman}
In this section, we provide our motivation and goals for both the Pommerman benchmark and the NIPS 2018 competition.
Currently, there is no consensus benchmark involving either general sum game settings or settings with at least three players. Instead, recent progress has focused on two player zero-sum games such as Go and Chess. We believe that the Pommerman environment can assume this role for multi-agent learning. Additionally, we are organizing competitions for Pommerman because we believe that they are a strong way to push forward the state of the art and can contribute to lasting results for years to come.


\subsection{Multi-Agent Learning}
Historically, a majority of multi-agent research has focused on zero-sum two player games. For example, computer competitions for Poker and Go over the past fifteen years have been vital for developing methods culminating in recent superhuman performance \cite{DBLP:journals/corr/MoravcikSBLMBDW17,ijcai2017-772,DBLP:journals/cacm/BowlingBJT17,SilverHuangEtAl16nature}. These benchmarks have also lead to the discovery of new algorithms and approaches like Monte Carlo Tree Search \cite{Vodopivec:2017:MCT:3207692.3207712,browne2012survey,Kocsis:2006:BBM:2091602.2091633,conf/cg/Coulom06} and Counterfactual Regret Minimization \cite{NIPS2007_3306}.

We believe that an aspect restraining the field from progressing towards general-sum research and scenarios with more than two players is the lack of suitable environments. We propose Pommerman as a solution.

Pommerman is stylistically similar to Bomberman \cite{bomberwiki}, the famous game from Nintendo. At a high level, there are at least four agents all traversing a grid world. Each agent's goal is to have their team be the last remaining. They can plant bombs that, upon expiration, destroy anything (but rigid walls) in their vicinity. It contains both adversarial and cooperative elements.  The Free-For-All (FFA) variant has at most one winner and, because there are four players, encourages research directions that can handle situations where the Nash payoffs are not all equivalent. 
The team variants encourage research with and without explicit communication channels, including scenarios where the agent has to cooperate with previously unseen teammates. The latter is a recently burgeoning subfield of multi-agent learning \cite{foerster2016learning,DBLP:journals/corr/abs-1804-07178,DBLP:journals/corr/EvtimovaDKC17,DBLP:journals/corr/FoersterNFTKW17,lewis2017deal,mordatch2017emergence,lazaridou2018emergence} with established prior work as well \cite{steels:99e,steels03,conf/eelc/LevyK06,fehervari:jr:10}, while the latter has been underexplored. 

We aim for the Pommerman benchmark to provide for multi-agent learning what the Atari Learning Environment \cite{Bellemare:2013:ALE:2566972.2566979} provided for single-agent reinforcement learning and ImageNet \cite{imagenet_cvpr09} for image recognition. Beyond game theory and communication, Pommerman can also serve as a testbed for research into reinforcement learning, planning, and opponent/teammate modeling.

RoboCup Soccer \cite{NardiNoda14} is a similar competition that has been running since 1997. There, eleven agents per side play soccer. Key differences between Pommerman and RoboCup Soccer are:
\begin{enumerate}
\item Pommerman includes an explicit communication channel. This changes the dynamics of the game and adds new research avenues. 
\item Pommerman strips away the sensor input, which means that the game is less apt for robotics but more apt for studying other aspects of AI, games, and strategy.
\item Pommerman uses low dimensional, discrete control and input representations instead of continuous ones. We believe this makes it easier to focus on the high level strategic aspects rather than low level mechanics.
\item In team variants, the default Pommerman setup has only two agents per side, which makes it more amenable to burgeoning fields like emergent communication which encounter training difficulties with larger numbers of agents.
\item Pommerman's FFA variant promotes research that does not reduce to a 1v1 game, which means that a lot of the theory underlying such games (like RoboCup Soccer) is not applicable.
\end{enumerate}

The second, third, and fourth differences above are a positive or negative trade-off depending on one's research goals.

Another, more recent, benchmark is Half-Field Offense \cite{ALA16-hausknecht}, a modification of RoboCup that reduces the complexity and focuses on decision-making in a simplified subtask. However, unlike the FFA scenario in Pommerman, Half-Field Offense is limited to being a zero-sum game between two teams.

In general, the communities that we want to attract to benchmark their algorithms have not gravitated towards RoboCup but instead have relied on a large number of one-off toy tasks. This is especially true for multi-agent Deep RL. We think that the reasons for that could be among the five above. Consequently, Pommerman has the potential to unite these communities, especially when considering that future versions can be expanded to more than four agents.

\begin{table*}[t]
\centering
{
\small
\begin{tabular}{|l|c|c|c|}
\thead{Game} & \thead{Intuitive?} & \thead{Fun?} & \thead{Integration?} \\ \hline
Bridge & 1 & 3 & 5 \\ \hline
Civilization & 2 & 3 & 1 \\ \hline
Counterstrike & 5 & 5 & 2 \\ \hline
Coup & 4 & 5 & 5 \\ \hline
Diplomacy & 1 & 4 & 3 \\ \hline
DoTA & 3 & 5 & 2 \\ \hline
Hanabi & 2 & 3 & 5 \\ \hline
Hearthstone & 1 & 4 & 1 \\ \hline
Mario Maker & 4 & 5 & 3 \\ \hline
\textbf{Pommerman} & \textbf{5} & \textbf{4} & \textbf{5} \\ \hline
PUBG & 5 & 5 & 1 \\ \hline
Rocket League & 5 & 4 & 1 \\ \hline
Secret Hitler & 4 & 4 & 3 \\ \hline
Settlers of Catan & 4 & 3 & 3 \\ \hline
Starcraft 2 & 3 & 5 & 5 \\ \hline
Super Smash & 5 & 5 & 1 \\ \hline
\end{tabular}
}
\caption{Comparing multi-agent games along three important axes for uptake beyond whether the game satisfies the community's intended research direction. Attributes are considered on a 1-5 scale where 5 represents the highest value. Fun takes into account both watching and playing the game. The Intuitive and Fun qualities, while subjective, are noted because they have historically been factors in whether a game is used in research.}
\label{table:game-comparisons}
\end{table*}

\subsection{High Quality Benchmark} 

There are attributes that are common to the best benchmarks beyond satisfying the community's research direction. These include having mechanics and gameplay that are intuitive for humans, being fun to play and watch, being easy to integrate into common research setups, and having a learning problem that is not \textit{too} difficult for the current state of method development. Most games violate at least one of these. For example, the popular game Defense of the Ancients \cite{dota} is intuitive and fun, but extremely difficult to integrate. On the other hand, the card game Bridge is easy to integrate, but it is not intuitive; the gameplay and mechanics are slow to learn and there is a steep learning curve to understanding strategy.

Pommerman satisfies these requirements. People have no trouble understanding basic strategy and mechanics. It is fun to play and to watch, having been developed by Nintendo for two decades. Additionally, we have purposefully made the state input based not on pixel observations but rather on a symbolic interpretation so that it does not require large amounts of compute to build learning agents.

Research game competitions disappear for two reasons - either the administrators stop running it or participants stop submitting entrants. This can be due to the game being `solved', but it could also be because the game just was not enjoyable or accessible enough. We view Pommerman as having a long life ahead of it. Beyond the surface hyperparameters like board size and number of walls, early forays suggest that there are many aspects of the game that can be modified to create a rich and long lasting research challenge and competition venue. These include partial observability of the board, playing with random teammates, communication among the agents, adding power-ups, and learning to play with human players. 

These potential extensions, and the fact that N-player learning by itself has few mathematical guarantees, suggest that Pommerman will be a challenging and fruitful testbed for years to come.

There are, however, limitations to this environment. One difficulty is that a local optimum arises where the agent avoids exploding itself by learning to never use the bomb action. In the long term, this is ineffective because the agent needs to use the bomb to destroy other agents. Players have successfully solved this challenge \cite{2018arXiv180706919R}, but it is an aspect of basic gameplay that has to be handled in order for the multi-agent research benefits to become apparent.

\section{Description}
In this section, we give details of the Pommerman environment. Note that all of the code to run the game and train agents can be found in our git repository \cite{github}, while our website (pommerman.com) contains further information on how to submit agents.

\subsection{Game Information}
\begin{figure}[h!]
    \centering
    \includegraphics[width=4cm]{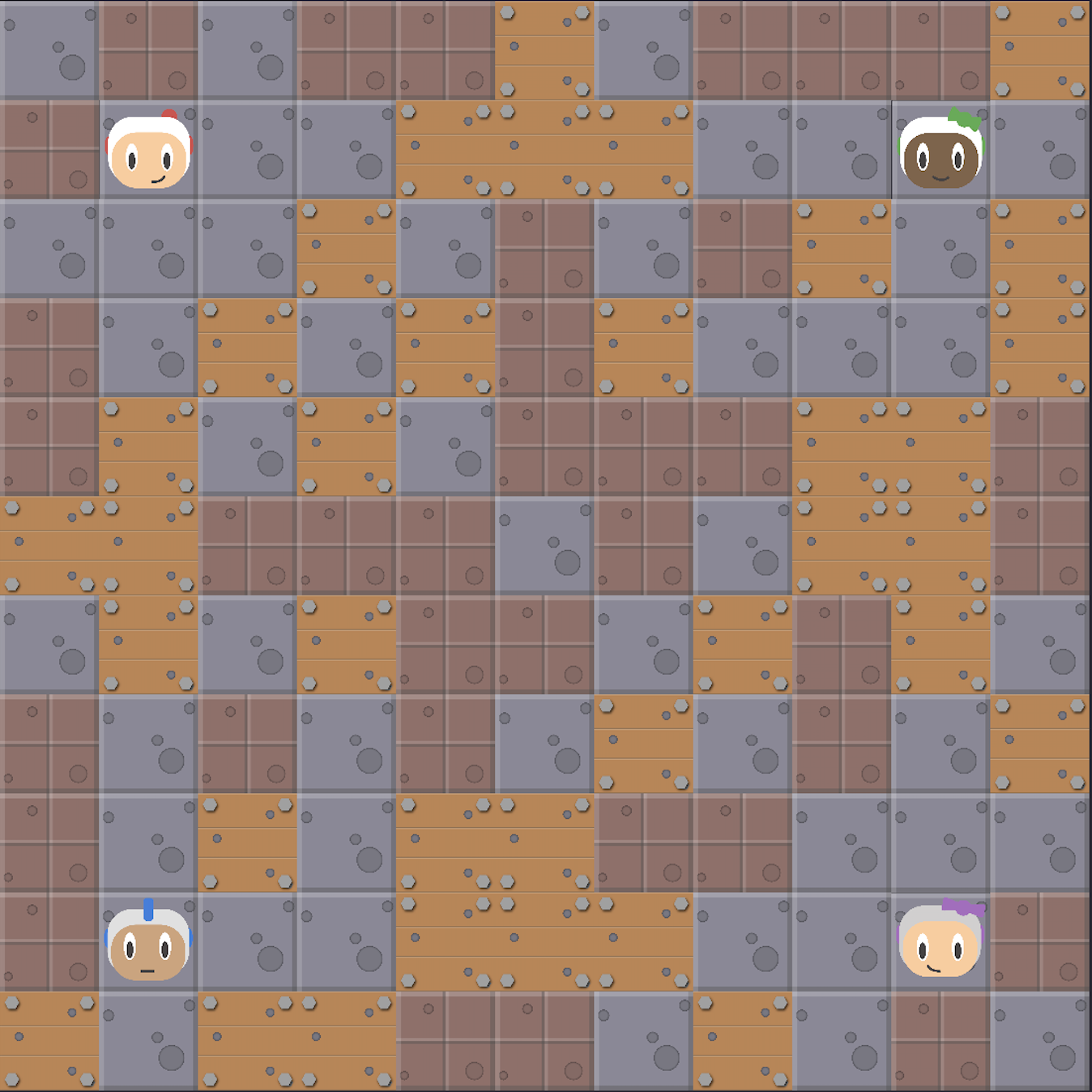}
    \caption{Pommerman start state. Each agent begins in one of four positions. Yellow squares are wood, brown are rigid, and the gray are passages.}
    \label{fig:pommerman-start}
\end{figure}

As previously mentioned, Pommerman is stylistically similar to Bomberman. Every battle starts on a randomly drawn symmetric 11x11 grid (`board') with four agents, one in each corner. Teammates start on opposite corners.

In team variants, the game ends when both players on one team have been destroyed. In FFA, it ends when at most one agent remains alive. The winning team is the one who has remaining members. Ties can happen when the game does not end before the max steps or if the last agents are destroyed on the same turn. If this happens in competitions, we will rerun the game. If it reoccurs, then we will rerun the game with collapsing walls until there is a winner. This is a variant where, after a fixed number of steps, the game board becomes smaller according to a specified cadence. We have a working example in the repository.

Besides the agents, the board consists of wooden and rigid walls. We guarantee that the agents will have an accessible path to each other. Initially, this path is occluded by wooden walls. See Figure \ref{fig:pommerman-start} for a visual reference.

Rigid walls are indestructible and impassable. Wooden walls can be destroyed by bombs. Until they are destroyed, they are impassable. After they are destroyed, they become either a passage or a power-up.

On every turn, agents choose from one of six actions:
\begin{enumerate}
    \item \emph{Stop}: This action is a pass.
    \item \emph{Up}: Move up on the board.
    \item \emph{Left}: Move left on the board.
    \item \emph{Down}: Move down on the board.
    \item \emph{Right}: Move right on the board.
    \item \emph{Bomb}: Lay a bomb.
\end{enumerate}

Additionally, if this is a communicative scenario, then the agent emits a message every turn consisting of two words from a dictionary of size eight. These words are passed to its teammate in the next step as part of the observation. In total, the agent receives the following observation each turn:
\begin{itemize}
	\item \emph{Board}: 121 Ints. The flattened board. In partially observed variants, all squares outside of the 5x5 purview around the agent's position will be covered with the value for fog (5). 
    \item \emph{Position}: 2 Ints, each in [0, 10]. The agent's (x, y) position in the grid.
    \item \emph{Ammo}: 1 Int. The agent's current ammo.
    \item \emph{Blast Strength}: 1 Int. The agent's current blast strength.
    \item \emph{Can Kick}: 1 Int, 0 or 1. Whether the agent can kick or not.
    \item \emph{Teammate}: 1 Int in [-1, 3].  Which agent is this agent's teammate. In non-team variants, this is -1.
    \item \emph{Enemies}: 3 Ints in [-1, 3]. Which agents are this agent's enemies. In team variants, the third int is -1.
    \item \emph{Bomb Blast Strength}: List of Ints. The bomb blast strengths for each of the bombs in the agent's purview. 
    \item \emph{Bomb Life}: List of Ints. The remaining life for each of the bombs in the agent's purview.
    \item \emph{Message}: 2 Ints in [0, 8]. The message being relayed from the teammate. Both Ints are zero only when a teammate is dead or if it is the first step. This field is not included for non-cheap talk variants.
\end{itemize}

The agent starts with one bomb (`ammo'). Every time it lays a bomb, its ammo decreases by one. After that bomb explodes, its ammo will increase by one. The agent also has a blast strength that starts at two. Every bomb it lays is imbued with the current blast strength, which is how far in the vertical and horizontal directions that bomb will effect. A bomb has a life of ten time steps. Upon expiration, the bomb explodes and any wooden walls, agents, power-ups or other bombs within reach of its blast strength are destroyed. Bombs destroyed in this manner chain their explosions.
        
Power-Ups: Half of the wooden walls have hidden power-ups that are revealed when the wall is destroyed. These are:
\begin{itemize}
    \item \emph{Extra Bomb}: Picking this up increases the agent's ammo by one.
    \item \emph{Increase Range}: Picking this up increases the agent's blast strength by one.
    \item \emph{Can Kick}: Picking this up permanently allows an agent to kick bombs by moving into them. The bombs travel in the direction that the agent was moving at one unit per time step until they are impeded either by a player, a bomb, or a wall.
\end{itemize}

\subsection{Early results}
The environment has been public since late February and the competitions were first announced in late March. In that time, we have seen a strong community gather around the game, with more than 500 people in the Discord server (https://discord.gg/mtW7kp) and more than half of the repository commits from open source contributors.

There have also been multiple published papers using Pommerman \cite{2018arXiv180706919R,zhou2018pommermanagent}. These demonstrate that the environment is challenging and we do not yet know what are the optimal solutions in any of the variants. In particular, the agents in \cite{2018arXiv180706919R} discover a novel way of playing where they treat the bombs as projectiles by laying, then kicking them at opponents. This is a strategy that not even novice humans attempt, yet the agents use it to achieve a high success rate.

Preliminary analysis suggests that the game can be very challenging for reinforcement learning algorithms out of the box. Without a very large batch size and a shaped reward \cite{ng1999policy}, neither of Deep Q-Learning \cite{mnih2013playing} nor Proximal Policy Optimization \cite{ppo} learned to successfully play the game against the default learning agent (`SimpleAgent'). One reason for this is because the game has a (previously mentioned) unique feature in that the bomb action is highly correlated with losing but must be wielded effectively to win.

We also tested the effectiveness of DAgger \cite{daume2009search} in bootstrapping agents to match the SimpleAgent. We found that, while somewhat sensitive to hyperparameter choices, it was nonetheless effective at yielding agents that could play at or above the FFA win rate of a single SimpleAgent ($\sim20\%$). This is less than chance because four simple agents will draw a large percentage of the time.

\section{Competitions}
In this section, we describe the Pommerman competitions. This includes both the upcoming NIPS 2018 event and the FFA competition that we already ran.

\subsection{FFA competition}
We ran a preliminary competition on June 3rd, 2018. We did not advertise this widely other than within our Discord social group (https://discord.gg/mtW7kp), nor did we have any prizes for it. Even so, we had a turnout of eight competitors who submitted working agents by the May 31st deadline.

The competition environment was the FFA variant \cite{pommermanffa} where four agents enter, all of whom are opponents. The top two agents were submitted by G{\"o}r{\"o}g Márton and a team led by Yichen Gong, with the latter being the strongest.

G{\"o}r{\"o}g's agent improved upon the repository's baseline agent through a number of edits. On the other hand, Yichen's agent was a redesign implementing a Finite State Machine Tree-Search approach \cite{zhou2018pommermanagent}. They respectively won 8 and 22 of their 35 matches (with a number of the remaining being ties).

\subsection{NIPS Competition}
The NIPS competition will be held live at NIPS 2018 and competitors are required to submit a team of two agents by November 21st, 2018. The featured environment will be the partially observable team variant without communication. Otherwise, we will be reusing the machinery that we developed to run the FFA competition.



\subsection{Submitting Agents}
We run the competitions using Docker and expect submissions to be accompanied by a Docker file that we can build on the game servers. For FFA competitions, this entails submitting a (possibly private) repository having one Docker file representing the agent. For team competitions, this means the submission should have two Docker files to represent the two agents. Instructions and an example for building Docker containers from trained agents can be found in our repository \cite{github}. 

The agents should follow the prescribed convention specified in our example code and expose an `act' endpoint that accepts the dictionary of observations. Because we are using Docker containers and http requests, we do not have any requirements for programming language or framework.

The expected response from the agent will be a single integer in [0, 5] representing which of the six actions that agent would like to take. In variants with messages, we also expect two more integers in [1, 8] representing the message. If an agent does not respond in an appropriate time limit for our competition constraints (100ms), then we will automatically issue them the Stop action and, if appropriate, have them send out the message (0, 0). This timeout is an aspect of the competition and not native to the game itself.

\section{Conclusion}
In this paper, we have introduced the Pommerman environment, detailed why it is a strong setup for multi-agent research, and described early results and competitions.

All of the code is readily available at our git repository (github.com/MultiAgentLearning/playground) and further information about competitions, including NIPS 2018, on our website (pommerman.com).

\section{Acknowledgments}
We are especially grateful to Roberta Raileanu, Sanyam Kapoor, Lucas Beyer, Stephan Uphoff, and the whole Pommerman Discord community for their contributions, as well as Jane Street, Facebook AI Research, Google Cloud, and NVidia Research for their sponsorship.

\bibliography{bibliography}
\bibliographystyle{aaai}
\end{document}